\documentclass{emulateapj}

\def\mics{$\mu$m}
\def\kms{km s$^{-1}$}
\def\ergs{erg s$^{-1}$}
\def\msun{M$_{\odot}$}
\def\smass{M$_{*}$}
\def\deluv{$\Delta(U-V)$}
\def\lfir{$L_{60}$}
\def\dellfir{$\Delta L_{60}$}

\slugcomment{To appear in Astrophysical Journal}

\shorttitle{Are AGN in quenching galaxies?}
\shortauthors{Rosario et al.}

\begin{document}

\title{Nuclear Activity is more prevalent in Star-Forming Galaxies}

\author{D.J. Rosario\altaffilmark{1}, P. Santini \altaffilmark{2}, D. Lutz\altaffilmark{1}, H. Netzer\altaffilmark{3}, 
F.E. Bauer\altaffilmark{4,10}, S. Berta\altaffilmark{1}, B. Magnelli\altaffilmark{5}, P. Popesso\altaffilmark{1}, 
D.M. Alexander\altaffilmark{6}, W.N. Brandt\altaffilmark{7}, R. Genzel\altaffilmark{1}, R. Maiolino\altaffilmark{8, 9}, 
J.R. Mullaney\altaffilmark{6}, R. Nordon\altaffilmark{3}, A. Saintonge\altaffilmark{1}, L. Tacconi\altaffilmark{1}, S. Wuyts\altaffilmark{1}} 

\altaffiltext{1}{Max-Planck-Institut f\"{u}r Extraterrestrische Physik (MPE), Postfach 1312, 85741 Garching, Germany; 
rosario@mpe.mpg.de, lutz@mpe.mpg.de, berta@mpe.mpg.de, popesso@mpe.mpg.de, 
genzel@mpe.mpg.de, amelie@mpe.mpg.de, linda@mpe.mpg.de, swuyts@mpe.mpg.de}
\altaffiltext{2}{ INAF - Osservatorio Astronomico di Roma, via di Frascati 33, 00040 Monte Porzio Catone, Italy;  paola.santini@oa-roma.inaf.it}
\altaffiltext{3}{School of Physics \& Astronomy, Tel Aviv University, 69978 Tel Aviv, Israel; netzer@wise.tau.ac.il, nordon@astro.tau.ac.il}
\altaffiltext{4}{Pontificia Universidad Cat\'{o}lica de Chile, Departamento de Astronom\'{\i}a y Astrof\'{\i}sica, Casilla 306, Santiago 22, Chile; fbauer@astro.puc.cl}
\altaffiltext{5}{Argelander-Institut f\"ur Astronomie, Auf dem H\"ugel 71, D-53121 Bonn, Germany;  magnelli@astro.uni-bonn.de}
\altaffiltext{6}{Department of Physics, Durham University, South Road, Durham, DH1 3LE, UK; d.m.alexander@dur.ac.uk, j.r.mullaney@durham.ac.uk}
\altaffiltext{7}{Department of Astronomy \& Astrophysics, 525 Davey Lab, The Pennsylvania State University, University Park, Pennsylvania 16802 USA; niel@astro.psu.edu}
\altaffiltext{8}{Kavli Institute for Cosmology, University of Cambridge, Madingley Road, Cambridge CB3 OHA, UK;  r.maiolino@mrao.cam.ac.uk} 
\altaffiltext{9}{Cavendish Laboratory, University of Cambridge, 19 JJ Thomson Avenue, Cambridge, CB3 0HE, UK}
\altaffiltext{10}{Space Science Institute, 4750 Walnut Street, Suite 205, Boulder, Colorado 80301}

\begin{abstract}

We explore the question of whether low and moderate luminosity Active Galactic Nuclei (AGNs) are
preferentially found in galaxies that are undergoing a transition from
active star formation to quiescence. This notion has been suggested by studies of the UV--optical colors
of AGN hosts, which find them to be common among galaxies in the so-called `Green Valley', a region of galaxy
color space believed to be composed mostly of galaxies undergoing star-formation quenching.
Combining the deepest current X-ray and Herschel/PACS far-infrared (FIR) observations
of the two Chandra Deep Fields (CDFs) with redshifts, stellar masses and rest-frame photometry derived 
from the extensive and uniform multi-wavelength data in these fields, we compare the rest-frame $U-V$ color distributions 
and SFR distributions of AGNs and carefully constructed samples of inactive control galaxies.
The UV-to-optical colors of AGNs are consistent with equally massive inactive galaxies at redshifts out to $z\sim2$,
but we show that such colors are poor tracers of star formation. While the FIR distributions of both
star-forming AGNs and star-forming inactive galaxies are statistically similar, we show that AGNs 
are preferentially found in star-forming host galaxies, or, in other words, AGNs are \emph{less} 
likely to be found in weakly star-forming or quenched galaxies. We postulate that, among X-ray selected
AGNs of low and moderate accretion luminosities, the supply of cold gas primarily determines
the accretion rate distribution of the nuclear black holes.
% \footnotemark[]
% \footnotetext[]{{\it Herschel} is an ESA space observatory with science instruments provided by European-led Principal Investigator consortia and with important participation from NASA.}
%, lending support to a dominant role of secular-mode fueling among such AGNs.

\end{abstract}


\section{Introduction}

Accreting black holes are arguably the most efficient engines of energy production in the Universe. The deep gravitational wells
of supermassive black holes (SMBHs) allow the extraction of $\sim10$\% of the rest mass-energy of the material that falls into
their horizons, which, through accretion processes, is ultimately converted into the electromagnetic and mechanical
output that power Active Galactic Nuclei (AGNs). This energy can, in turn, escape into the environs
of the AGN host galaxy, affecting material on large scales. Such AGN `feedback' has many
potential effects on galaxy physics and evolution: regulation of the circum-nuclear environment and galactic star-formation (SF),
gaseous outflows, the distribution of metals, enrichment and heating of gas in the circum-galactic medium.
There is much observational evidence for the direct effects of SMBH activity on their host galaxies, either
in the form of AGN-driven outflows 
\citep{holt08, alexander10, feruglio10, fischer10, greene11, rupke11, sturm11, cano12, harrison12a, maiolino12} 
or bubbles blown in the host atmospheres of massive galaxy haloes
by powerful radio jets \citep{mcnamara07,gitti12}.

Proposed evidence for the widespread action of SMBH feedback on the star-formation histories of galaxies
comes from suggestions that most AGNs occupy a preferred population of host galaxies, those that are undergoing
a transformation from active steady star-formation to a final state of quiescence \citep[e.g.,][]{nandra07, martin07,schawinski10}. 
These two phases form the basis
of the well known color-bimodality of galaxies -- most galaxies from the local Universe to $z\sim3$ lie in two relatively
distinct parts of the color-magnitude or color-mass diagram, the Blue Cloud and the Red Sequence. Galaxies that
lie at intermediate colors on these diagrams are believed to be transitioning between the two populations, through
a region known as the `Green Valley' \citep[e.g.,][]{faber07}. The preponderance of Green Valley galaxies among 
AGN hosts has been taken as evidence that low-to-moderate luminosity AGNs are responsible for the quenching of 
star-formation in galaxies, mediated through feedback from the SMBH. 
The closely associated morphological evolution of transforming galaxies \citep{driver06, kauffmann06, franx08, cheung12} is taken to be evidence that a substantial 
fraction of AGNs are related to, and perhaps triggered by, galaxy mergers which may be responsible for the formation of 
`Red and Dead' ellipticals.

In recent work, the notion that AGN prefer quenching hosts has come under greater scrutiny. The importance of stellar
mass selection effects for the interpretation of AGN colors seems to suggest that AGN host colors are not radically 
different from similarly massive inactive galaxies \citep{xue10, rosario11}. A previous study by our team \citep{santini12}
found that mean SFRs are enhanced in X-ray AGN over inactive galaxies of the same stellar mass, with tentative
evidence that the enhancement was caused by a lower fraction of quiescent galaxies among AGN hosts, rather than a boost
in the SFRs of star-forming AGNs. However, given the relatively shallow far-infrared data used in this earlier work, this
notion could not be tested extensively.

In this paper, we critically examine the evidence that AGN are preferentially in galaxies that are quenching
or undergoing a slow-down of their global star-formation. Rather than relying only on optical or UV tracers of star-formation
such as the color-mass diagram (CMD), we employ instead the most sensitive far-Infrared (FIR)
data currently available from the Herschel Space Telescope, together with a large sample of AGNs selected from the deepest
extragalactic X-ray surveys on the sky -- the two Chandra Deep Fields. In \S2, we present the surveys, selection and
datasets. In \S3, we compare the use of FIR tracers and UV--optical colors in studies of SF. In \S4 and 5, we re-examine
the evidence that AGNs lie preferentially in the Green Valley, then fold in information about the FIR luminosities
and detection rates of AGNs toward exploring the question of whether AGNs are in quenching galaxies. Our results
are discussed in \S6. In this work, we assume a $\Lambda$-CDM concordance cosmology with $\Omega_{\Lambda}=0.7$, 
$\Omega_{m}=0.3$ and $H_0 = 70$ \kms Mpc$^{-1}$.

\section{Datasets and Sample Selection}

%%%%%%%%%%%%%%%%%%%%% FIGURE 1 %%%%%%%%%%%%%%%%%%%%%%%%
\begin{figure*}[t]
\figurenum{1}
\label{cmd_fir}
\centering
\includegraphics[width=\textwidth]{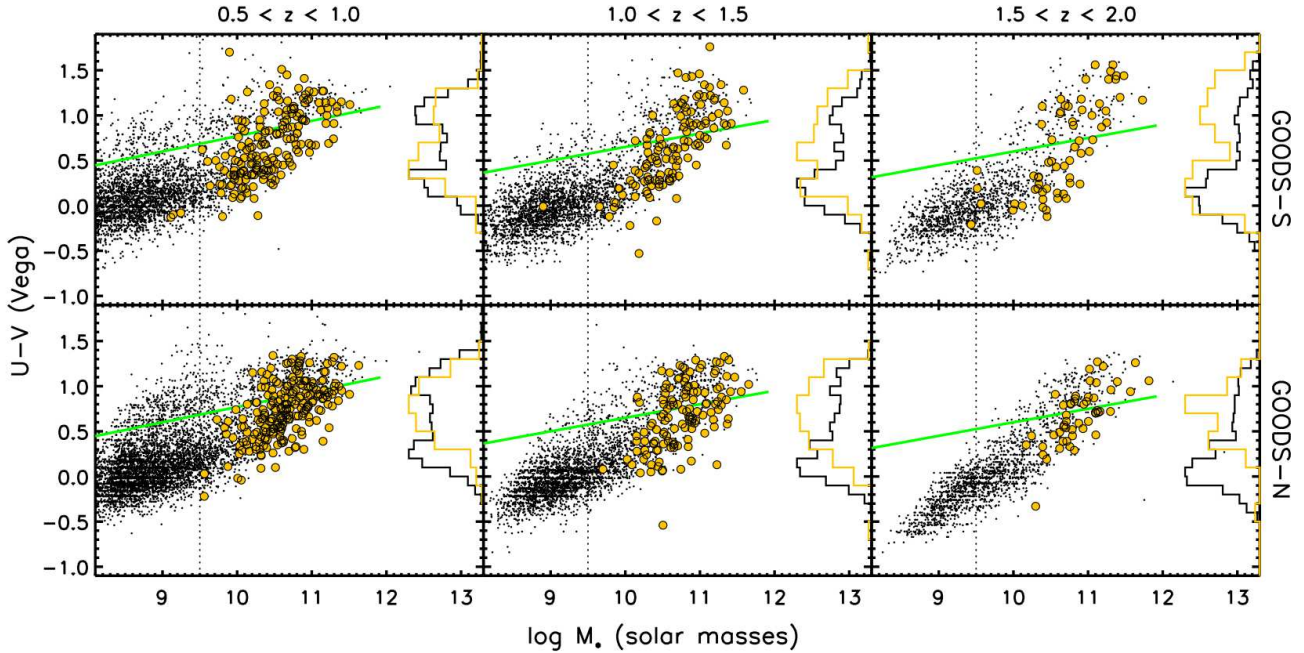}
\caption[CMD of galaxies with FIR]
{ Color-mass Diagrams (CMDs) of rest-frame $U-V$ color against stellar mass (\smass). All X-ray sources and IRAC-selected AGNs
have been excluded from these diagrams. The general galaxy population
taken from the full galaxy catalogs in both GOODS fields are shown as small black points, with the distribution of $U-V$
color for galaxies with \smass$>10^{9.5}$\msun\ (dotted line) shown as a black histogram on the right axis of each panel. A color bimodality
is seen at all redshifts. The location of the `Green Valley' is shown by solid green lines which mark 
the locus of minimum galaxy density in this diagram (see \S3 and Table 1). FIR-detected galaxies (for a definition, see \S2) 
are shown as large colored points and their $U-V$ distribution is shown on the right axis of each panel as a colored histogram.
Both histograms are normalized to have the same peak value. 
Despite lying on or above the SF Mass Sequence, most FIR detected galaxies exhibit $U-V$ colors that
are intermediate between the Red and Blue sequences, i.e, they lie in the Green Valley and show a weak
color bimodality. This is primarily because of their high stellar masses.
}
\end{figure*}
%%%%%%%%%%%%%%%%%%%%%%%%%%%%%%%%%%%%%%%%%%%%%%%%%

\subsection{Sample Selection and Datasets}

Cospatial with the two GOODS survey fields \citep{giavalisco04}, the Chandra Deep Fields (CDFs) are the deepest pencil-beam X-ray surveys
in the sky. In GOODS-North, the CDF-North (CDF-N)
X-ray catalog comprises 503 sources from a total exposure of 2 Msec  \citep{alexander03}, while in GOODS-South, the new
4 Msec CDF-South (CDF-S) X-ray catalog consists of 740 sources \citep{xue11}. We have extensively characterized the data and 
catalogs in both fields, in which careful associations have been made with optical and near-IR counterparts,
using, where possible, probabilistic crossmatching models \citep{luo10, xue11}. In addition to 
the deep X-ray data, the wealth of deep spectroscopy and multi-wavelength photometric data in the GOODS fields
have enabled accurate spectroscopic or AGN-optimised photometric
redshifts to be determined for the majority of the X-ray sources \citep[e.g.,][]{szokoly04, luo10}. We estimate absorption-corrected
hard-band X-ray luminosities ($L_X$) of sources with redshifts using spectral modeling techniques \citep{bauer04}.
As a result of the small area and great depth of the CDF exposures, most X-ray sources are low or moderate
luminosity AGNs -- only $\sim 5$\%\ of the sources have $\log L_{X} \textrm{(2-10 keV)} > 44$ \ergs. These have equivalent
AGN bolometric luminosities to the Seyfert galaxy population found in the local Universe.
In this work, we only consider sources with $\log L_{X} > 42$ \ergs, to prevent contamination from powerful starbursts, in which
emission from X-ray binaries can potentially overpower the emission from nuclear activity in such faint systems.

We employ multiwavelength galaxy catalogs for the two GOODS fields to define a general galaxy sample, the
properties of which we will compare to the AGNs. In GOODS-S, we use the updated 
GOODS-MUSIC database \citep{santini09, grazian06}, while in GOODS-N we use a catalog 
developed for the PEP team using similar methodologies \citep{berta10, berta11}
\footnote{Available on the \anchor{http://www.mpe.mpg.de/ir/Research/PEP/public_data_releases.php}{PEP public release page}.}. 
The former catalog selects galaxies 
with observed magnitudes in the HST F850LP band $<26$ or in the ISAAC $K_s$ band $<23.5$, while the 
latter is primarily selected to have $K<24.2$. In order to exclude a surfeit of faint sources with inaccurately red
colors and masses, we apply an additional cut of F850LP $<26$ in the GOODS-N catalog. For galaxies with no current
spectroscopic redshifts, photometric redshifts were determined by fitting multiwavelength photometry using
PEGASE 2.0 templates \citep{fioc97} in GOODS-S or using the EAZY code \citep{brammer08} 
in GOODS-N. For details on the catalog preparation, characterization and photometric redshift estimation, we
refer the reader to \cite{santini09} and \cite{berta10} for GOODS-S and GOODS-N respectively.

While AGN are selected by their X-ray emission, we define our `inactive' galaxy population as all galaxies that are
undetected in X-rays (excluding even those which have  $\log L_{X} < 42$ \ergs) and with mid-IR (MIR) colors (i.e, based
on Spitzer/IRAC photometry) that do not satisfy the AGN selection criteria of \cite{donley12}. In practice, only a very small
fraction of the general galaxy population are rejected on the basis of these criteria. These rejected objects, however, tend
to be in massive galaxies and could potentially sway the statistics of SF comparisons among such systems 
by an inordinate degree.

We have developed a custom technique for the estimation of stellar masses (\smass) in AGNs 
by linearly combining galaxy population synthesis model templates and AGN SED templates to fit multiwavelength
photometry. For inactive galaxies, we perform a
$\chi^{2}$ minimization of \cite{bc03} synthetic models, assuming a Salpeter Initial Mass Function (IMF) and parameterizing
the star formation histories as exponentially declining laws. For AGNs, we also include
an AGN template from \cite{silva04}, which accounts for a variable fraction of the total light of the galaxy. The AGN template
reflects the classification of the X-ray source, derived from information about its SED and
spectrum, where available. For sources classified as Type I (broad lines in the spectrum, clear AGN contribution
in the rest-frame optical and UV), a Seyfert 1 SED was used, while for the rest, a Seyfert 2 template was used
if the estimated X-ray absorption column $N_H < 10^{24}$ cm$^{-2}$, and a Compton-thick template for
more heavily absorbed systems. For further details, performance evaluations and tests of the method,
we refer the reader to \cite{santini12}.

%%%%%%%%%%%%%% Table 1 %%%%%%%%%%%%%%%%%%
\begin{deluxetable}{ccc}
\tablenum{1} 
\tablewidth{0pt}
%\tablewidth{\columnwidth}
\tablecaption{Parameters of the Green Valley (GV) lines in the Color-Mass Diagram}
\label{gvtable}
\tablehead{\colhead{Redshift interval} & \colhead{$UV_{9}$} & \colhead{$\alpha$}}
\startdata
0.5--1.0 &0.60 &0.17 \\
1.0--1.5 &0.50 & 0.15 \\
1.5--2.0 &0.45 & 0.15 \\
\enddata
\tablecomments{$UV_{9}$ and $\alpha$ are defined such that $(U-V)_{GV} = \alpha \log(M_{*,9}) + UV_{9}$, 
where $(U-V)_{GV}$ is the color of the GV line and $M_{*,9}$ is the stellar mass in units of $10^9$ solar masses.}
\end{deluxetable}
%%%%%%%%%%%%%%%%%%%%%%%%%%%%%%%%%%%%%

Our far-infrared data are composed of maps  at 70 \mics, 100 \mics\  and 160 \mics\ from a 
combination of two large Herschel/PACS programs: the PACS Evolutionary Probe (PEP), a guaranteed time program \citep{lutz11} 
and the GOODS-Herschel key program \citep{elbaz11}. The combined PEP+GH (PEP/GOODS-Herschel) 
reductions are described in detail in \citet{magnelli12}. 
While data at 100 and 160 \mics\ are available in both fields, an additional deep map at 70 \mics\ is also available in GOODS-S.
The PACS 160, 100 and 70 \mics\ fluxes were extracted using sources from archival 
deep Spitzer MIPS 24 \mics\ catalogs as priors, following the method described in \citet{magnelli09}; see also \citet{lutz11}
for more details. 3$\sigma$ depths are 0.90/0.54/1.29 mJy at 70/100/160 \mics\ in the central region of 
GOODS-S and 0.93/2.04 mJy at 100/160 \mics\ in GOODS-N. The GOODS-S maps are $\approx 80$\%\ deeper 
than the GOODS-N maps and probe further down the FIR luminosity function at all redshifts \citep{magnelli12}.

For practical purposes, we use the monochromatic luminosity of a galaxy at 60 \mics\ rest (\lfir) 
as a measure of its FIR luminosity. The PACS bands cover this rest-frame wavelength over much of the redshift
range probed in this work and we estimate \lfir\ from a simple log-linear interpolation of PACS measurements
in bands that bracket 60 \mics\ in the rest-frame. The use of \lfir\ obviates the need to apply an uncertain
correction between monochromatic and total FIR luminosities. Nevertheless, in order to relate \lfir\ to the properties of the
population of star-forming galaxies from the existing literature, such as the SF `Mass Sequence' or `Main Sequence' (MS), we adopt the
following relationship between \lfir\ and SFR:

\begin{eqnarray}
L_{IR} & = & f_{CE01} \times L_{60} \\
SFR & = & 1.72 \times 10^{-10} L_{IR}
\end{eqnarray}

$f_{CE01}$ is the conversion factor between \lfir\ and the total IR luminosity integrated over 8-1000\mics\ ($L_{IR}$) 
from \citet{ce01}.  This factor is IR luminosity dependent, but only varies slightly over the range $1.6$--$2.5$. The conversion
factor between SFR and $L_{IR}$ is taken from \citet{kennicutt98}, which also assumes a Saltpeter IMF.

%%%%%%%%%%%%%%%%%%%%% FIGURE 2 %%%%%%%%%%%%%%%%%%%%%%%%
\begin{figure*}[t]
\figurenum{2}
\label{ms_duv}
\centering
\includegraphics[width=\textwidth]{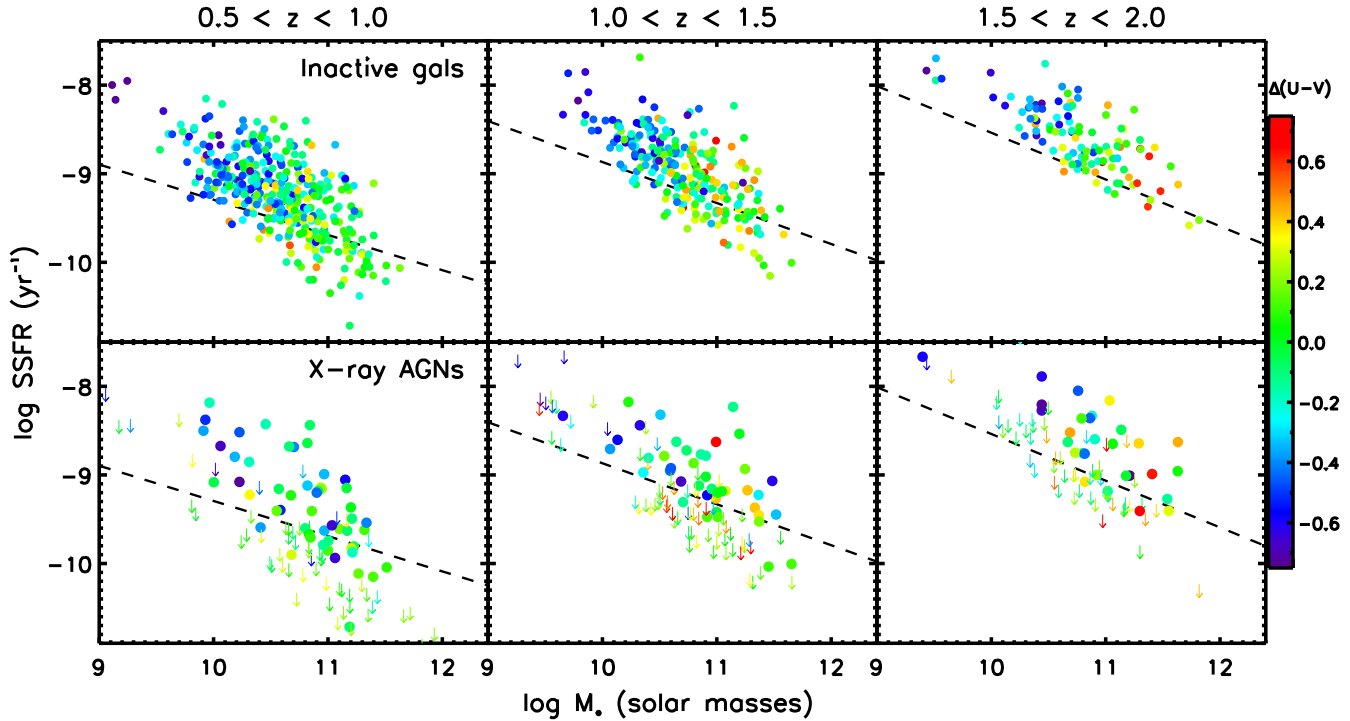}
\caption[MS colored by del UV]
{ Specfic Star Formation Rate (SSFR) against stellar mass (\smass) of PACS-detected inactive galaxies (top panels) and X-ray
selected AGNs (bottom panels). For AGNs undetected in PACS, we also show estimated upper limits on SSFR (arrow points). 
Points are colored by the \deluv, the $U-V$ offset of the galaxy from the Green Valley (GV) as defined by straight lines in the CMD (\S3).
Galaxies that lie in the GV have \deluv\ close to zero and are colored as green points in the Figure.
Both PACS-detected galaxies and PACS-detected AGNs in the GV lie on or around the Mass Sequence 
(shown as dashed lines) and are typically massive, while those in the Blue Cloud are typically at lower 
masses and lie well above the Mass Sequence. A substantial fraction of objects in the GV are normal massive SF galaxies at 
all redshifts in this study. The distribution of both PACS-detected AGNs and inactive galaxies about the Mass Sequence are 
formally indistinguishable, as developed further in Section 5.
}
\end{figure*}
%%%%%%%%%%%%%%%%%%%%%%%%%%%%%%%%%%%%%%%%%%%%%%%%%

\section{The colors and masses of FIR-detected inactive galaxies}

%%%%%%%%%%%%%% Table 1 %%%%%%%%%%%%%%%%%%
%\begin{table}[t]
%\tablenum{1}
%\label{gvtable}
%\begin{center}
%\begin{tabular}{|c|c|c|}
%\hline  \hline
%Redshift interval & $UV_{9}$ & $\alpha$ \\
%\hline
%0.5--1.0 &0.60 &0.17 \\
%1.0--1.5 &0.50 & 0.15 \\
%1.5--2.0 &0.45 & 0.15 \\
%\hline \hline
%\end{tabular}
%\end{center}
%\caption{Parameters of the Green Valley (GV) lines in the U-V vs. \smass\ plane. $UV_{9}$ and $\alpha$ are defined 
%such that $U-V$(GV) = $UV_{9} \log(M_{*}/10^{9}) + \alpha$, where $U-V$(GV) is the color of GV line and $M_{*}$
%is stellar mass in solar masses.}
%\end{table}
%%%%%%%%%%%%%%%%%%%%%%%%%%%%%%%%%%%%%

We begin by first examining the relationship between the optical color of galaxies and the star-formation rate
or star-formation luminosity, as traced by FIR emission. For this, we turn to the color-mass diagram (CMD)
as a diagnostic tool. In Fig.~\ref{cmd_fir}, we plot the rest-frame $U-V$ color against the stellar
mass \smass\ of galaxies, separately for the two GOODS fields. Consistent with several previous studies
in these and other extragalactic fields \citep[e.g.,][]{willmer06, wyder07, taylor09, brammer09, whitaker11}, 
galaxies tend to cluster in two well-defined regions of the diagram: the Red
Sequence and the Blue Cloud. These structures form the basis of the well known color-bimodality of galaxies, which
has been revealed to $z>2$ \citep{brammer09}. In such optical- or NIR-selected photometric catalogs, the mass
limit, revealed by the sharp boundary in the density of galaxies at the lower mass end of the diagram, is redshift
and color dependent. At $z>1.5$, the mass limit is high enough among red galaxies that it blurs the definition
of the color bimodality. In addition, at these redshifts, a population of extremely dust-reddened galaxies is also seen,
leading to a tail of very red colors among high mass galaxies. 

In between the dense Red Sequence and Blue Cloud, galaxies have intermediate colors, which
has led to the popular name for this area of the CMD: the Green Valley (GV). Taking the GV as a minimum in the
density distribution of galaxies on the CMD at a given stellar mass, we construct sloped lines on the CMD which
separate the Red Sequence from the Blue Cloud and define the location of the GV for  all our subsequent analysis. 
The slope and normalization of the lines were determined by eye to yield the most well-defined separation between the
Red Sequence and Blue Cloud in each redshift bin. We tabulate the GV lines for each bin in Table \ref{gvtable}
and plot them in Fig.~\ref{cmd_fir} using solid green lines.

It has been suggested, based on the low density of the GV, that most galaxies here are going 
through a relatively rapid phase of SF quenching, resulting in a net flow of galaxies from
the SF Blue Cloud to the quiescent Red Sequence \citep{faber07,martin07,brammer11}. In other 
words, GV galaxies are believed to have lower specific SFRs compared to normal SF galaxies, 
which are defined to lie on the SF Mass Sequence \citep{noeske07,elbaz07,santini09,
rodighiero11,wuyts11,whitaker12}.

Fig.~\ref{cmd_fir} suggests otherwise. We have plotted in this Figure, using large colored symbols, the locations of PACS
detected inactive galaxies from the PEP+GH catalogs. In general, these galaxies lie on or above the MS
at $z>0.5$ and tend to be rather massive (\smass$\gtrsim10^{10}$ \msun). It is clear from the plot that a substantial, 
number of FIR-bright galaxies lie in the GV. In fact, the main determinant of whether or not
a galaxy is detected in PACS is its stellar mass, not its color. 

This can be examined in a different way using Fig.~\ref{ms_duv}, where we have plotted the FIR-derived specific SFR (SSFR) of
PACS-detected inactive galaxies against \smass (upper panels). The ridgeline of the SF MS at the central redshift of each redshift bin
is shown in these plots as dashed lines, as determined recently by \citet{whitaker12}.
The points here are colored by \deluv, the $U-V$ color offset of the galaxy from the GV in the CMD, i.e the vertical offset
of the galaxy from the green GV lines in Fig.~\ref{cmd_fir}. 

The steeper slope of the PACS-detected galaxies compared to the MS lines is due to the flux limit of the PACS photometry,
which translates into a mass-dependent limit in SSFR at a given redshift. At lower \smass, only galaxies that
lie at progressively higher SFR above the MS are detectable in PEP+GH. This strongly affects the interpretation
of the slope of the MS purely from FIR data, which is why we adopt MS relationships determined, in this case, from
a uniform study of deeper UV and 24\mics\ photometry \citep{whitaker12}. We have changed the mass scaling in 
the relations of \citet{whitaker12}, estimated using a Chabrier IMF, upwards by a factor of 1.74 to reflect our use of the Salpeter IMF
\citep{santini12a}. 

At \smass$>10^{10.5}$ \msun, we see that galaxies that lie around the 
MS have colors that are typical of the GV, i.e with \deluv$\approx 0$ \citep[see also][]{whitaker12, salmi12}. 
In essence, this means that massive galaxies with intermediate colors are not a special population that is
quenching and moving away from a state of on-going SF, but, indeed, \emph{are} the star-forming 
population at \smass$\gtrsim 10^{10.5}$ \msun.

In later sections, we examine the PACS detection fractions of AGNs and inactive galaxies. It is worthwhile, at this stage,
to note that PACS detected galaxies primarily lie on or above the MS, even though the MS is known to have a roughly 
symmetrical scatter of about 0.3 dex in SSFR \citep{noeske07, rodighiero11}. Concentrating only on PACS detected galaxies 
restricts us to galaxies that scatter above the MS, but one must remember that there are normal star-forming 
galaxies that nominally lie within the MS but are below our detection limit since they lie within the scatter below the MS.
When discussing PACS detection statistics, we lump together quiescent and quenching galaxies, 
which are expected to lie well below the normal scatter of the MS, and `weakly SF galaxies', those that lie in the low 
scatter of the MS. The reader should keep in mind that the population of PACS-undetected galaxies are not all quiescent
or quenching, but will in addition include a sizable number of actively SF galaxies that are simply below our nominal PACS 
detection limits.

\section{The colors of X-ray AGN: Comparison to Inactive Galaxies}

%%%%%%%%%%%%%%%%%%%%% FIGURE 3 %%%%%%%%%%%%%%%%%%%%%%%%
\begin{figure*}[t]
\figurenum{3}
\label{cmd_agn}
\centering
\includegraphics[width=\textwidth]{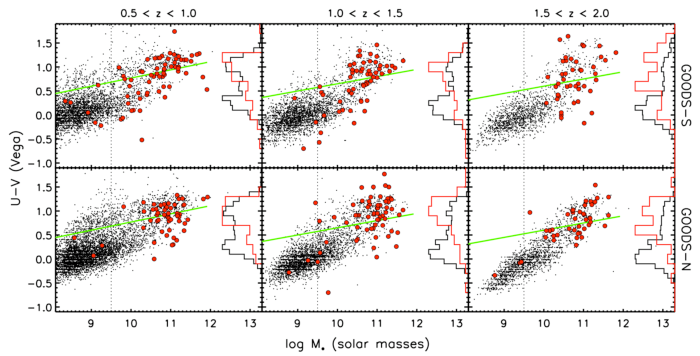}
\caption[CMD of galaxies and AGN]
{ Color-mass Diagrams (CMDs) of rest-frame $U-V$ color against stellar mass (\smass) comparing AGNs and inactive galaxies. 
The general galaxy population of inactive galaxies are shown as small black points, with the distribution of $U-V$
color for galaxies with \smass$>10^{9.5}$\msun\ (dotted line) shown as a black histogram on the right axis of each panel.  
X-ray selected AGNs are shown as large colored points and their $U-V$ distribution is shown on 
the right axis of each panel as a colored histogram.
Both histograms are normalized to have the same peak value. The location of the `Green Valley' is shown by solid green lines
as in Fig.~\ref{cmd_fir}. AGNs exhibit typically 
high stellar masses, similar to FIR-detected star-forming galaxies (compare this plot to Fig.~\ref{cmd_fir}).
}
\end{figure*}
%%%%%%%%%%%%%%%%%%%%%%%%%%%%%%%%%%%%%%%%%%%%%%%%%

We have shown the rest-frame optical color is not an ideal tracer to identify a quenching population
because its use as a measure of the SF properties of a galaxy depends on the stellar mass of the galaxy.
Nevertheless, several previous studies have taken the location of AGNs in the CMD as evidence for
a link between AGN activity and the transformation of galaxies \citep[e.g.,][]{nandra07,martin07,schawinski10}. 
The reason for this can be seen in Fig.~\ref{cmd_agn}.
The AGNs, shown as large colored points, scatter mostly about the GV lines that we defined in \S3, with a small
scatter to lower masses and colors in the Blue Cloud. In many ways, the approximate distribution of AGN in the CMD
mirrors that of PACS-detected inactive galaxies. However, this is almost completely because both AGNs and FIR-bright
galaxies are typically quite massive. 

Indeed, it is this strong tendency for X-ray detected AGNs to lie in massive host galaxies that mostly determines
the location of the AGNs in the CMD. Recent studies suggest that, once the particular mass distribution of
AGN hosts is taken into account, the colors of AGNs and inactive galaxies are very similar \citep{silverman09, xue10, 
cardamone10, rosario11}. 
The small sizes of AGN samples, generally only a few \%\ of galaxies, make it difficult to identify 
statistically robust differences in the colors of AGNs and inactive galaxies. Here we develop a method that builds
on the much larger sample of inactive galaxies to test the following null hypothesis: AGN hosts 
are drawn randomly from the population of massive galaxies and share the SFRs and colors
of the parent sample.

For each AGN, we choose, at random and allowing duplicates, an inactive galaxy in the same redshift bin and with a stellar 
mass within $\pm 0.1$ dex of the mass of the AGN host galaxy. In this way, we arrive at a sample of
inactive galaxies which are equal in number and mass distribution as the AGNs in the redshift bin. 
The mass tolerance is smaller than the typical error in stellar masses from our SED fits, ensuring essentially
identical mass distributions of AGNs and inactive galaxies. 
From this comparison sample, we derive a distribution in color. We repeat this process for a total
of 1000 trials. This bootstrap approach gives us the typical uncertainty in the 
color distribution for inactive galaxies that share the mass distribution of AGNs.

%%%%%%%%%%%%%%%%%%%%% FIGURE 4 %%%%%%%%%%%%%%%%%%%%%%%%
\begin{figure*}[t]
\figurenum{4}
\label{duv_xagn}
\centering
\includegraphics[width=\textwidth]{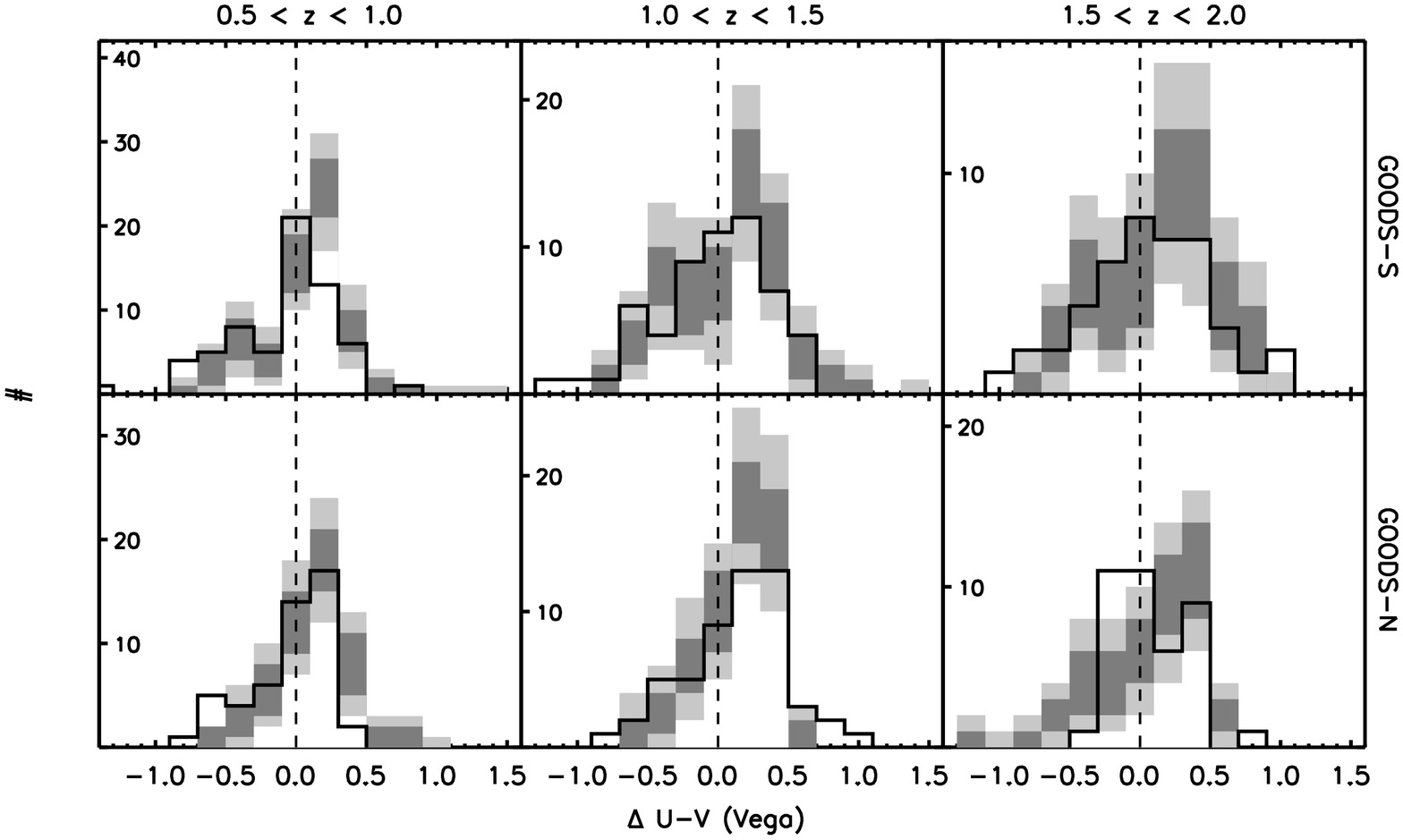}
\caption[GV color offset for AGN]
{A comparison of the Green Valley color offset (\deluv) of X-ray selected AGNs and mass-matched
inactive galaxies. The open histograms drawn with a solid black line show the distribution of \deluv\ for the
AGNs, while shaded histograms correspond to the inactive galaxies. 
The statistical uncertainty in the distributions of the inactive galaxies are shown
by the shading in the histograms - dark grey sections show the 1$\sigma$ uncertainty, due to
the scatter in the population as well as small number statistics, while the light grey sections
show 2$\sigma$. The dashed line at \deluv$=0$ is the location of the center of the Green Valley.
The AGNs show very similar distributions to inactive galaxies, especially among the bluer star-forming
population, but are under-represented in the Red Sequence.}
\end{figure*}
%%%%%%%%%%%%%%%%%%%%%%%%%%%%%%%%%%%%%%%%%%%%%%%%%

In Fig.~\ref{duv_xagn}, we compare \deluv, the $U-V$ color offset from the Green Valley, of AGNs and
inactive galaxies. The distribution of \deluv\ for the AGNs are shown as a open histogram, plotted
over the distributions of an equal number of inactive galaxies with the same redshift and stellar mass range as the AGNs. 
Using our bootstrap procedure, we obtain $1\sigma$ and $2\sigma$ uncertainties on the inactive galaxy distributions,
shown as dark and light grey shaded regions in the Figure. The colors of AGNs are statistically different
from those of the inactive galaxies only if the open histogram strongly or consistently deviates from the darkly 
shaded regions over a few or more bins in \deluv. Note that systematic differences in the distributions
due to cosmic variance are not represented in the uncertainties of these shaded histograms, though they can
be important in small fields like GOODS and among the massive galaxies considered here. A comparison using
both fields helps in this regard.
 
A quick examination shows that, in almost all redshift bins, there are very minor differences between AGNs and inactive galaxies.
At the location of the Green Valley, the AGNs are slightly more common than inactive galaxies, at the level of 1-1.5$\sigma$
in five of six panels, but this difference is quite small and may be attributable to enhanced line emission or recombination 
continuum emission in AGNs, since nuclear activity can frequently produce extended, highly ionized, emission line regions.
At about the same low level of significance, there is an enhancement of inactive galaxies over AGNs in the Red Sequence
(\deluv\ in the range $0.2-0.6$ mag). Perhaps the most significant difference between the two distributions is not near
their peak, but is in the extreme blue wing, where AGNs are more common than inactive galaxies. This is likely due to the small
fraction (a few \%) of AGNs which contain bright blue nuclear point sources which can dominate the integrated light of the system.
These will have very blue colors for their stellar mass, a result of nuclear contamination in the optical and UV bands.

\section{The FIR-derived SFRs of X-ray AGN: Comparison to Inactive Galaxies}

%%%%%%%%%%%%%%%%%%%%% FIGURE 5 %%%%%%%%%%%%%%%%%%%%%%%%
\begin{figure}[t]
\figurenum{5}
\label{agn_cont}
\centering
\includegraphics[width=\columnwidth]{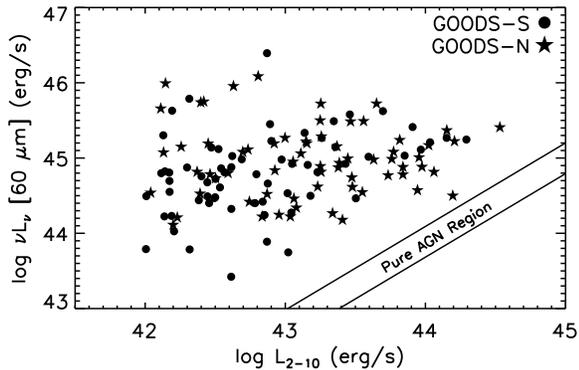}
\caption[AGN contamination in FIR]
{\lfir\ plotted against the X-ray luminosity of the AGNs in both Chandra Deep Fields over the redshift range $0.5<z<2.0$.
Different symbols are used to represent sources from the two different survey fields. The region between the two
solid lines is where one may expect to find AGN with negligible star-formation and FIR SEDs dominated by dust
heated by the nucleus (see \S5 and \citep{mullaney11a}). Essentially, all the AGN have FIR luminosities well above the 
pure-AGN region. It is highly unlikely that AGN emission powers the FIR luminosity of these sources, even among those at the
luminous end. Note, the apparent weak correlation between $L_{2-10}$ and \lfir\ is a consequence of Eddington bias and
is not a real trend.}
\end{figure}
%%%%%%%%%%%%%%%%%%%%%%%%%%%%%%%%%%%%%%%%%%%%%%%%%

%%%%%%%%%%%%%%%%%%%%% FIGURE 6 %%%%%%%%%%%%%%%%%%%%%%%%
\begin{figure*}[t]
\figurenum{6}
\label{irccds}
\centering
\includegraphics[width=\textwidth]{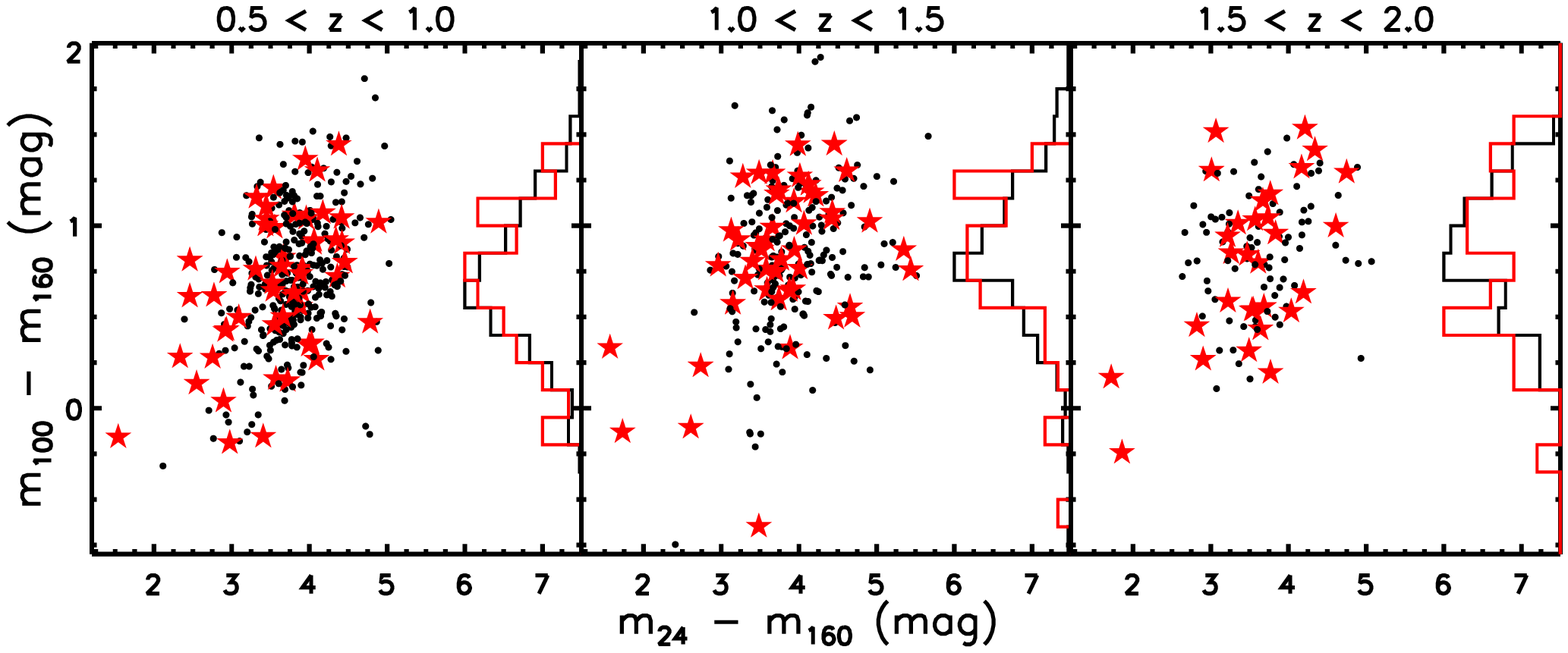}
\caption[IR colors of AGN and galaxies]
{Infrared colors (in magnitudes) of PACS-detected AGNs (red star points) and inactive galaxies (small black points) 
from both GOODS fields combined. 24 -- 160 \mics\ (a MIR-FIR color) is plotted on the X-axis and
100 -- 160 \mics\ (a FIR-FIR color) is plotted on the Y-axis. Distributions of 100 -- 160 \mics\ colors are shown
as histograms on the Y-axes, with red/black colors for AGNs/inactive galaxies respectively. Both histograms
are normalized to the same peak value. There is a small scatter of AGNs to low, blue MIR-FIR colors, due to the
influence of AGN-heated dust on the 24 \mics\ emission, but most AGNs lie in the area of the diagram occupied
by inactive SF galaxies, implying that their FIR emission, if not the entire MIR-to-FIR SED, is dominated
by SF-heated dust emission. 
}
\end{figure*}
%%%%%%%%%%%%%%%%%%%%%%%%%%%%%%%%%%%%%%%%%%%%%%%%%

We now turn to a far better tracer of SF in galaxies:  the FIR luminosity. Among massive galaxies, such as X-ray selected
AGN hosts, the FIR luminosity, an excellent tracer of the total dust-reprocessed UV light from star-forming regions, accounts
for almost all the SF in the galaxy -- the UV escape fractions in such galaxies is quite low and corrections for un-reprocessed
SF luminosity are at the level of a few \%\ \citep{pannella09, reddy10, whitaker12}. 

The relative SF properties of AGN hosts and inactive galaxies can be understood more accurately
by comparing their FIR luminosity distributions. Studies of the FIR SEDs of AGNs show that the contribution
to \lfir\ from dust emission heated by the active nucleus is low and generally negligible, except in a few cases of luminous
AGNs in weakly SF galaxies \citep{netzer07, mullaney11a, rosario12a}. We can directly verify this for FIR-bright AGNs in the
CDF fields, as done in Fig.~\ref{agn_cont}, where we compare measured \lfir\ of AGNs detected in PACS with the 
predicted FIR luminosity derived from $L_X$ based on model relationships of pure AGN-heated dust. 
These relationships, described in \citet{rosario12a}, have SED shapes covering the scatter
found among local X-ray bright AGNs \citep{mullaney11a}, while the nuclear X-ray luminosity
is linked to the IR luminosity following the tight correlation of \cite{gandhi09}. From the Figure, it is clear
that the AGN luminosities of most of the sources in our sample are too low to substantially affect the FIR luminosity.

Another simple test is to compare the IR colors of AGNs and inactive galaxies; if the FIR colors of AGNs are significantly
warmer than inactive galaxies (i.e, with higher relative flux in bluer bands), then AGN heating may be responsible for
a non-negligible fraction of \lfir. In Fig.~\ref{irccds}, we plot the MIPS 24 \mics\ to PACS 160 \mics\ color against 
the PACS 100 \mics\ to 160 \mics\ color (both expressed as magnitudes) of PACS detected AGNs and inactive galaxies. 
This plot may be directly compared to Fig.~2 of \cite{rosario12a}, which studied the IR colors of sources in the COSMOS
survey in similar fashion. The FIR color distributions of AGNs and inactive galaxies (histograms along the Y-axis in all panels) are 
formally indistinguishable, though, expectedly, AGN emission can influence the MIR, as evidenced by the scatter of
AGN points to bluer 24-to-160 \mics\ colors. For essentially all of our PACS-detected AGNs, we can safely 
assume that \lfir\ is a measure of the SFR. 

\begin{figure*}[t]
\figurenum{7}
\label{dlfir_xagn}
\centering
\includegraphics[width=\textwidth]{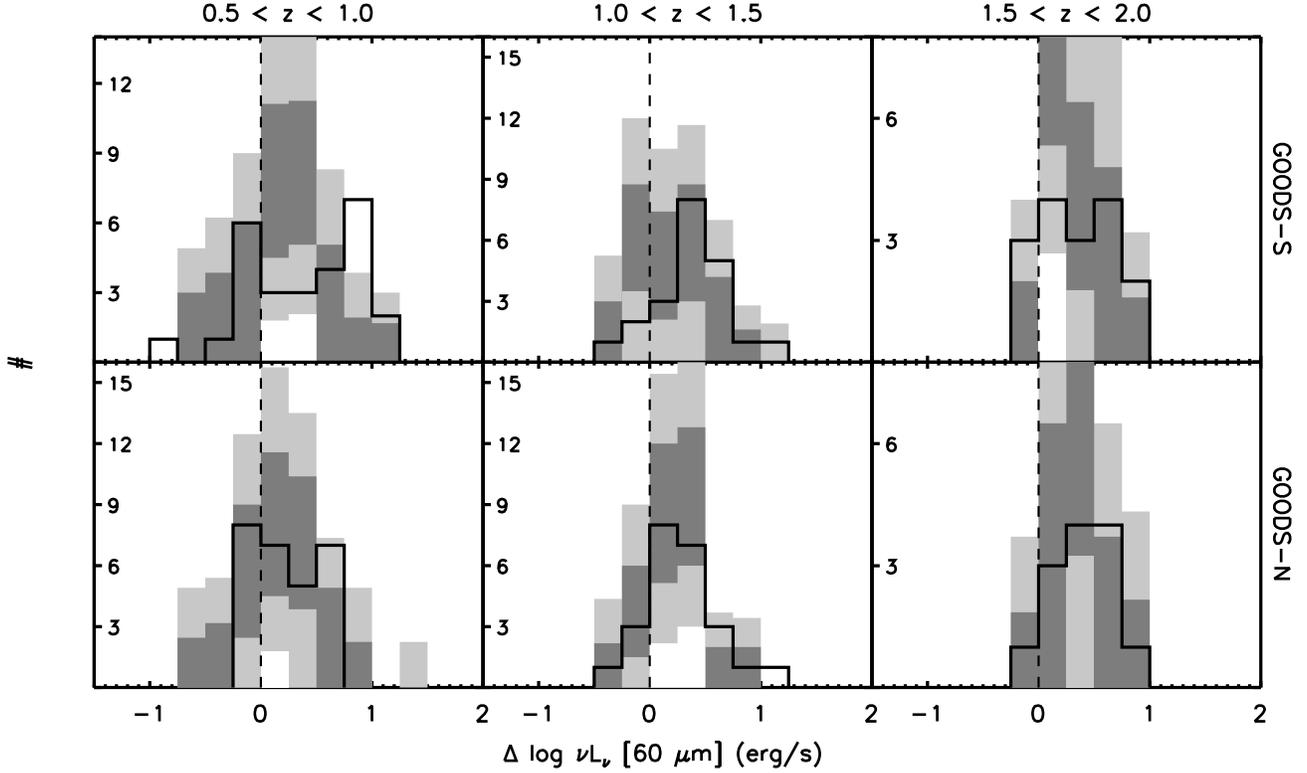}
\caption[MS Offset for AGN]
{A comparison of the \lfir\ offset (\dellfir) of X-ray selected AGNs and mass-matched
inactive galaxies from the star-formation Mass Sequence. The statistical uncertainty in the distributions of the inactive galaxies are shown
by the shading in the histograms - dark grey sections show the 1$\sigma$ uncertainty, due to
the scatter in the population as well as small number statistics, while the light grey sections
show 2$\sigma$. The dashed line at \dellfir $=0$ corresponds to the center of the Mass Sequence.
The AGNs show rather similar distributions to inactive galaxies.}
\end{figure*}

Using FIR-based SFRs, AGNs share a similar scatter about the Mass Sequence 
ridgeline with inactive galaxies (lower panels of Fig.~\ref{ms_duv}), including
a number of AGNs with GV colors that lie squarely on the Mass Sequence. We can evaluate this more rigorously by comparing
\dellfir, the offset of a galaxy from the \citet{whitaker12} Mass Sequence, for AGNs and mass-matched
inactive galaxies (Fig.~\ref{dlfir_xagn}). The known redshift and stellar mass of each object determines the corresponding MS value; then
the \cite{ce01} IR SED libraries and the relationship from \citet{kennicutt98} are used to connect the MS to \lfir\ (\S2). 
In practice, the choice of comparing simple \lfir\ distributions or \dellfir\ distributions does not alter our basic result; we use the
offset from the MS simply to scale out any stellar-mass dependent trend which would otherwise broaden the distributions. 
As for Fig~\ref{duv_xagn}, we use a bootstrap sampling procedure, randomly selecting a set of inactive galaxies matched to the AGNs in
\smass, repeated a thousand times, to arrive at statistical uncertainties in the \lfir\ distributions of the inactive galaxies. Dark/light grey
regions show $1\sigma$ and $2\sigma$ uncertainties respectively. 

Unlike in \deluv, there is a great deal of scatter in the inactive galaxy population in terms of FIR luminosity, as evinced by
$2\sigma$ uncertainties, which can span from zero to almost twice the number of AGNs at some values of \dellfir. Despite this, we find that
the AGN distributions are generally consistent with the inactive galaxy distributions at the 1-1.5$\sigma$ level, i.e, the solid histograms
lie within or close to the dark-grey regions. There is a possible minor tendency for an excess of PACS-detected
SF galaxies at or below the \dellfir$=0$ line compared to AGNs in the GOODS-S panels at $z>1$. 
However, this is not seen in GOODS-N and may just be due to cosmic variance. The broad consistency between the distributions
implies that any differences in SF properties between AGN hosts and inactive galaxies that lie on or
above the SF Mass sequence are minor or non-existent, a result that is completely consistent with earlier work 
on the CDF X-ray selected AGN based on independent PEP and GOODS-Herschel data \citep{mullaney11b, santini12}
or from studies from Spitzer/MIPS FIR surveys \citep[e.g.,][]{juneau13}. Star-forming AGN hosts are 
drawn from the general population of SF galaxies.

Galaxies with the greatest SF offset from the MS are known to be strong starbursts and frequently show highly
disturbed morphologies consistent with being galaxy mergers \citep{wuyts11}. 
The \dellfir\ distribution is dominated by galaxies close to the ridgeline
of the MS and we have rather poor statistics for AGN and galaxies at the highest \dellfir\ ($\gtrsim 10$), 
especially considering the substantial scatter shown in the distributions of inactive galaxies. Nevertheless, it is clear that
AGNs are not strongly enhanced among starbursting galaxies, at least among the range of 
nuclear luminosities probed in the work. 

%%%%%%%%%%%%%%%%%%%%% FIGURE 7 %%%%%%%%%%%%%%%%%%%%%%%%
%%%%%%%%%%%%%%%%%%%%%%%%%%%%%%%%%%%%%%%%%%%%%%%%%

There are, however, some clear differences between the populations of AGN hosts and inactive galaxies which come to light
once we consider sources that are not detected in the PACS maps. These sources typically lie below the MS and
include quenched galaxies and weakly star-forming galaxies (i.e, normal SF galaxies that lie within the lower 
scatter of the MS). To understand this population, we compute the FIR `non-detection' fraction, $f_{nd}$
which is the fraction of galaxies in each redshift bin that have no flux in both 100 and 160 \mics\
PACS maps at the $3\sigma$ detection threshold of the catalogs. $f_{nd}$ is a measure of the relative number of weakly 
star-forming or quiescent galaxies in a population.

\begin{figure*}[t]
\figurenum{8}
\label{undetfracs}
\centering
\includegraphics[width=\textwidth]{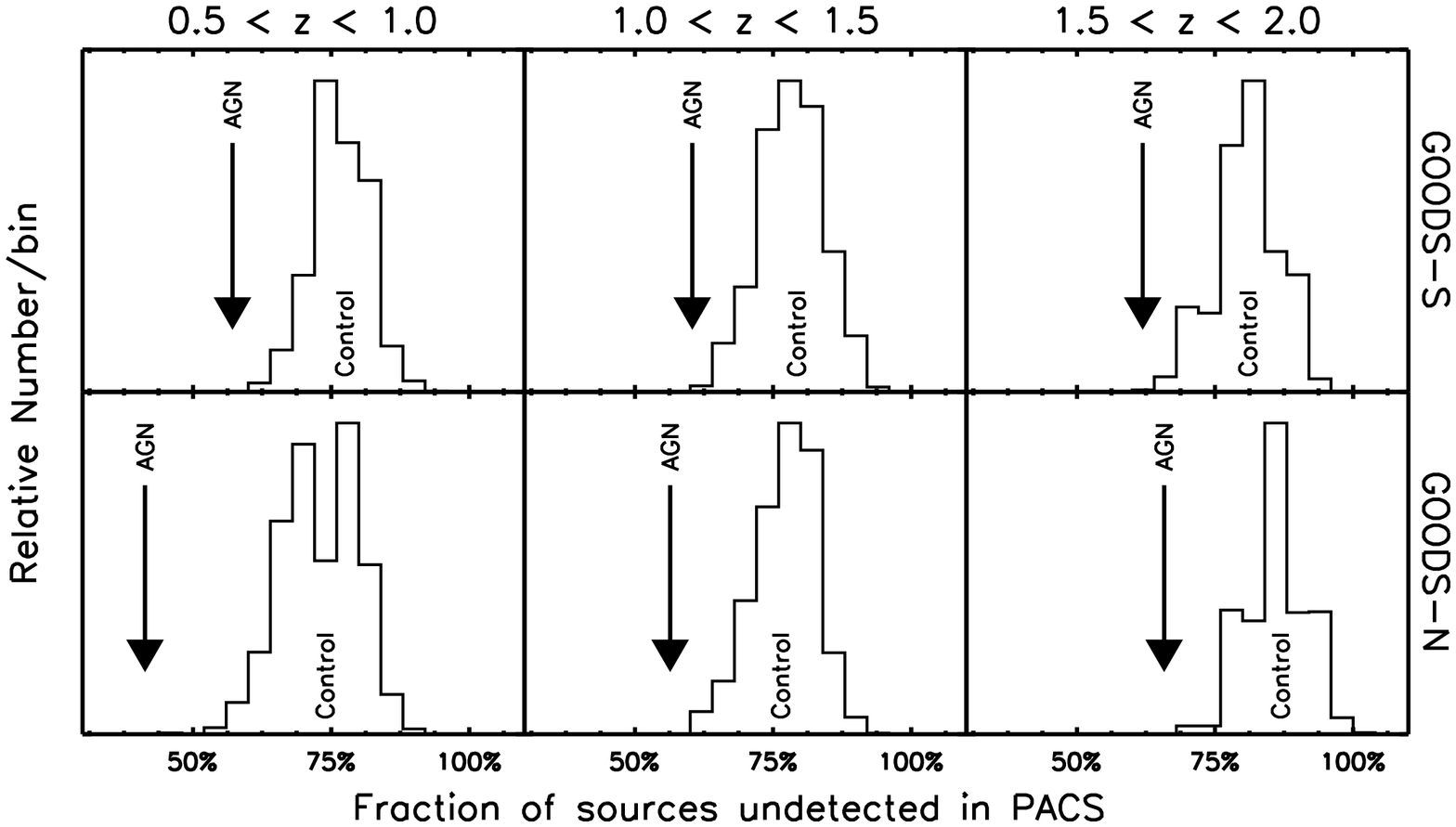}
\caption[Non-detection fractions for AGN]
{Histograms of the PACS non-detection fractions -- the percentage of objects \emph{not} detected in PEP+GH
PACS maps -- for 1000 realizations of the mass-matched comparison sample of inactive galaxies in the
corresponding redshift bins. The median value of the histograms is shown by the location of the vertical label `Control'.
The non-detection fraction of X-ray AGNs in the same redshift bin is shown as a thick
arrow for comparison. AGNs have a significantly higher chance of being detected in the deep Herschel data, which
implies, given the depth of the PEP+GH maps, that they preferentially avoid 
weakly star-forming, quenching or quiescent galaxies.}
\end{figure*}

In Fig.~\ref{undetfracs}, we plot the histogram of $f_{nd}$ for inactive galaxies
determined from the 1000 realizations of the mass-matching bootstrap procedure described above. The median non-detection
fractions for this population are approximately $80$\%. This fraction is roughly independent of redshift despite the evolving
luminosity limit of the PACS data, which reflects the fact that galaxies at $z\sim2$ are typically
more star-forming, and hence FIR luminous, than at $z\sim0.5$. In these insets, we show the value of $f_{nd}$ for the AGNs
in that redshift bin, a single number, using a thick downward-pointing arrow. It is immediately clear that the AGNs are always,
at all redshifts and consistently in both fields, significantly more likely to be detected in PACS than the inactive galaxies. In other
words, the fraction of quenching or quiescent galaxies hosting AGNs is considerably lower than similarly massive inactive galaxies,
or, put differently, AGN are more likely to be in star-forming galaxies around or above the MS. This result is in stark
contrast to the notion that AGN are preferentially in quenching systems.  Controlling for the particular mass distribution
of X-ray selected AGN hosts, we show that, in fact, AGN are \emph{more likely} to be found in a galaxy that is 
forming stars, compared to one that is turning quiescent. This result strongly confirms the result from \citet{santini12}
that AGN have higher FIR detection rates, which leads to the enhancement in mean SFR among AGNs compared
to normal galaxies.

%%%%%%%%%%%%%%%%%%%%% FIGURE 8 %%%%%%%%%%%%%%%%%%%%%%%%
%%%%%%%%%%%%%%%%%%%%%%%%%%%%%%%%%%%%%%%%%%%%%%%%%

\subsection{Biases and tests}

We have undertaken a series of tests to verify that this result is not due to possible biases in the data. Since the CDF sensitivity
has a radial variation, the density of X-ray sources is higher in the central area of the field, while the comparison sample
of background galaxies has a more uniform spatial distribution. Therefore, a possible bias may arise if 
the equivalent PACS coverage, which varies across the PEP+GH maps, differs 
between the AGNs and inactive sample. An examination of the PACS data coverage of AGNs and inactive 
galaxies shows that they are quite consistent; this is because the parts of the GOODS
fields with strong PACS coverage gradients are restricted to the edges. We then repeated our analysis of FIR properties, 
while matching the AGNs and control galaxies to within a factor of 1.5 in PACS coverage, in addition to \smass. 
This severely limits the number of matched comparison galaxies for some of the AGNs in our sample, but does 
not change our main conclusions. Any differences in the spatial distribution of the AGNs and the comparison galaxy 
sample on the sky does not lead to a lower $f_{nd}$ of the AGNs. 

Another possible source of bias arises from the use of MIR (MIPS 24\mics) sources as priors during the photometric extraction
of the PACS maps. In the small number of cases where a PACS source was a blend of multiple MIPS sources,
FIR fluxes were decomposed using the 24 \mics\ fluxes as weights. Since AGNs can be more MIR bright
than inactive galaxies as a result of warm dust emission from the circumnuclear regions and torus,
they could be assigned a stronger weight in blended sources, which can bias the FIR fluxes of AGN, as a population,
to inaccurately high values. To test this, we repeated our analysis but only including AGNs that had no 
neighboring 24 \mics\ MIPS sources within a radius of 5". The results were unchanged, implying that extraction biases, if
present, have a negligible effect on the FIR photometry of X-ray selected AGNs in the CDFs. We conclude that the low $f_{nd}$ of AGNs 
is robust to systematic and data-related selection effects. 

Finally, we consider the effects of AGN-heated dust emission on the FIR detection rates of AGNs.
We have shown that the majority of these AGNs have luminosities that are too low to boost a 
significant fraction of weakly star-forming hosts into the PACS detectable regime -- if this were 
the explanation of the low quenching fractions, PACS-detected AGNs in the CDFs should generally show
very warm/blue FIR colors, which is not observed (Fig.~\ref{irccds}). 

\section{Discussion: Are AGN hosts a special population of galaxies?}

We have undertaken a detailed analysis of the FIR luminosities and detection rates of low and moderate
luminosity AGNs in the Chandra Deep Fields in the context of the population of inactive galaxies. 
Our key finding is the high FIR detection rates of AGNs, implying that they are significantly less likely to be hosted by
galaxies undergoing a special phase of star-formation quenching, contrary to the results of earlier studies.
This difference arises because, at the high masses of most AGN host galaxies, the UV-to-optical color is a poor proxy for 
SFR, which has complicated earlier studies of the SF properties of AGNs based on the use of color-mass or color-magnitude
diagrams. Our PACS detection rates for AGNs ($\sim 50$\%) are higher than those reported in the
Spitzer/MIPS 70 \mics\ studies of \citep{juneau13}, but this is due to the significantly deeper PEP+GH data.

We also find that low and moderate luminosity X-ray AGNs have similar distributions of FIR 
luminosities as inactive SF galaxies, consistent with the notion that host galaxies of such AGNs are
drawn from a population of mostly normal galaxies. In a study of $z\sim0$ emission-line
selected AGNs from the SDSS, \citet{pasquali05} find evidence for a small (0.2 dex) enhancement in mean 
FIR luminosities over a control sample of normal SF galaxies, matched to the AGNs
by multiple structural and photometric criteria. Given the difference in the AGN selection method and matching scheme,
we are unable to make a detailed comment on the difference between our results and those from \citet{pasquali05}. However,
a small offset of the magnitude uncovered in that work may still be consistent with our findings, given the
substantial uncertainties on the \dellfir\ distributions in Fig.~\ref{dlfir_xagn} and the fact that the Figure excludes 
some weakly star forming galaxies. Alternatively, a study of UV-based SFRs of SDSS Type II AGNs by \citet{salim07}
find that local ``high-luminosity'' AGNs, which are actually comparable in nuclear luminosity to the X-ray selected AGNs from this work,
lie on the local Mass Sequence, consistent with our result. 

What could lead to the peculiar SF nature of AGN hosts? To answer this, we need to consider the relative timescales
of bright AGN activity (i.e, detectable in X-ray surveys of  high redshift galaxies) and the modulation of SF
on galaxy scales. It is highly unlikely that SMBHs accrete at a constantly high rate to maintain detectable AGN activity for
a long period of time; typical constraints on the lifetime of a bright QSO phase suggest $10^{7-8}$ years \citep{martini01, 
hopkins05a,shankar10}.
Even during a period of substantial accretion, the luminous output of an AGN may change quite significantly. High resolution 
simulations of SMBH accretion, both with and without the effects of feedback, suggest that 
the supply of gas to the accretion disk is not expected to flow in at a constant and steady pace \citep{hopkins10,novak11}. Therefore,
the X-ray bright AGN population is probably a rather transient population among galaxies in the field.

On the other hand, SF on scales of the entire galaxy varies on timescales comparable to its dynamical time of around $10^{8}$ yr.
In addition, the FIR luminosity, arising mostly from dust heated by stars of a range of ages, is not a prompt measure of
SFR, but can average over timescales of tens to hundreds of Myr, especially in systems with fairly steady SF histories. 

In light of this, we may consider two possible alternatives to explain our observations:

A) Some evolutionary models suggest that a major fraction of moderate luminosity AGNs arise in the aftermath
of a strong starburst, during which the SMBH also grows in a high accretion-rate QSO-like phase \citep[e.g.,][]{ciotti07,hopkins08a}. 
A popular form of these
models suggests that gas-rich major mergers modulate both starbursts and QSOs. The end product is a quenching
post-starburst galaxy hosting a low-to-moderate luminosity AGN as the final amounts of gas fall intermittently into the SMBH. 
Within this scenario, the high PACS detection rates of AGN hosts may plausibly be ascribed to the gradual decline 
in dust heating by a post-starburst population, which serves as a fossil record of the co-evolutionary phase which occurred within
the last few 100 Myrs. This picture would require a fine-tuning of the timescale between the peak of SF and the
period over which low and moderate luminosity AGNs exist, in order to place the AGNs on the SF Mass Sequence
with the same distribution of SSFRs as smoothly evolving inactive galaxies. It is also inconsistent with the observation
that the [O II]-derived SFRs of moderate luminosity AGNs are comparable to normal star-forming galaxies \citep{silverman09}: emission
line tracers are a more prompt measure of SF than the FIR and the starburst co-evolution scenario would therefore predict a lower
SF-powered line luminosity in AGNs than most comparable inactive galaxies. These observations disfavor the
post-starburst/major merger picture as the driver for most low and moderate luminosity AGN activity.

There are suggestions that, among heavily obscured X-ray undetected AGNs at similar redshifts, the SFR is enhanced
compared to the Mass Sequence \citep[e.g.,][]{juneau13}. One of the difficulties inherent in making the comparison of our
work with such studies is the very different selection functions of X-ray selected AGNs and, for e.g, emission-line or
MIR-selected AGNs. 
%In addition, the level of contamination from inactive SF galaxies in the selection of obscured AGN samples is not yet fully characterized.
Taking a recent example, \citet{juneau13} find that the fraction of AGNs in a 70 \mics\ detected sample of galaxies
remains constant across SSFR, consistent with our result, but only if one includes Mass-Excitation (MEx) 
selected objects undetected in the X-rays. However, the typical stellar mass of the MEx-selected AGNs in 
that study is significantly lower than the X-ray detected AGNs, which automatically gives such objects a 
higher SSFR distribution in their FIR luminosity-limited sample (for a demonstration of this, notice
the high SSFRs of low mass AGNs in the lower panels of Fig.~\ref{ms_duv}).  For a fair comparison between our
work and studies of obscured AGNs, full account of both redshift and stellar mass dependent selection effects
needs to be taken. We aim to address obscured AGNs in future PEP+GH work following the methodology developed here.

B) A second view postulates that AGN hosts are simply drawn from a smoothly evolving population
of massive star-forming galaxies. The connection here between star-formation and AGN activity is through
the supply of cold gas needed for both, mediated through evolutionary processes that modulate
SF in galaxy disks and carry gas to the nucleus for eventual accretion onto the SMBH. The lack of any
clear relationship between nuclear and SF luminosities in low and moderate luminosity AGNs 
\citep{shao10, mullaney11b, rosario12a, harrison12b}
argues against a prompt or synchronized connection between galaxy and black hole growth, favoring this
scenario. The lower incidence of AGN among weakly star-forming or quiescent galaxies is a consequence
of the depleted supply of cold gas in such systems; SMBH activity in such galaxies is primarily in the form of
a mechanically-dominated low accretion rate mode, such as that found in most radio-loud AGNs \citep{churazov05,trump11,best12}.
%Indeed, this may argue for a more critical role for radio-mode feedback in hastening the quenching of galaxies,
%rather than photon feedback from optical and X-ray bright phases. 

If cold gas is necessary for fueling X-ray bright AGN, then how do we reconcile this with the existence of such AGNs
among genuinely quiescent galaxies, albeit at a low rate? This is indeed a puzzle, but is likely related to the fact that cold gas is not 
entirely absent in such systems. As many as 22\%\ of nearby early-type galaxies harbor substantial molecular gas \citep{young11}
and some show clear SF \citep{crocker11}. Dusty disks with very low SF efficiencies are commonly seen in the circumnuclear
regions of massive elliptical galaxies, while filamentary dust is frequent on larger scales \citep{lauer95,ferrarese06}.
\cite{kauffmann09} have suggested that SMBH accretion rates in quiescent galaxies are set by the supply of gas
from stellar mass loss. Indeed, only a small quantity of this gas needs to intermittently reach the nucleus to fuel
such low-luminosity AGNs. 

Existing surveys of large, complete samples of X-ray selected AGNs suggests the existence of a universal accretion rate
distribution for SMBHs, which is independent of the stellar mass of the host \citep{aird12, bongiorno12}. Our results can qualify these results
by noting that the accretion rate distribution of AGN hosts do vary with SFR - galaxies with on-going star-formation are
more likely to host a modestly accreting black hole than those that are quiescent. At lower stellar masses,
both the gas supply and the SMBH mass drops, which is what possibly maintains the universal accretion rate
distribution. However, the relative number of SF to quenched galaxies increases greatly with redshift--- by $z\sim2$,
70\%\ of massive galaxies are widely forming stars \citep{fontana09}. Concurrent to this is an increase in the gas fractions 
of galaxies \citep{daddi10, tacconi10, tacconi12}. Therefore, the accretion rate distribution must change with redshift, both 
in normalization and shape, if, indeed, the supply of gas is what limits the time-averaged accretion rate of SMBHs. 
The form of this change would be a shift to a higher characteristic break in the accretion rate distribution, 
as more SMBHs of a given mass accrete from the more abundant cold gas supply at high redshifts. This is consistent with \citep{mullaney12} who show that average BH growth rates in mass-matched samples of galaxies indeed increase with redshift in line with their SFRs.

As a final point, we consider the effects of galaxy dust properties, which could alter the interpretation of our results.
If AGN are preferentially in galaxies with larger dust masses, this would lead to redder UV--optical colors
over galaxies with smaller dust masses or lower average extinction. However, since dust
is closely tied to gas, large dust masses would imply large gas reservoirs and, generally high SFRs, which only
serves to underline our findings. The difficulty of using $U-V$ (or other similar colors) to understand SF properties in
massive galaxies is compounded by the effects of dust extinction on these colors. For example,
\citet{cardamone10} find that AGN hosts are common among galaxies where dust reddening has pushed colors from the star-forming Blue
Cloud to the Green Valley. This qualitatively agrees with our findings that AGNs are predominantly in star-forming
galaxies. \citet{cardamone10} also suggest that AGN hosts in the Red Sequence and in star-forming galaxies
have different accretion histories and feedback mechanisms. This may be true, but we maintain that, given
the presence of molecular dusty gas even in genuinely quiescent galaxies, cold gas accretion may still be responsible
for some of the AGN activity on the Red Sequence.

\section{Summary}

We characterize and study the star-forming properties of X-ray selected active galaxies in the Chandra Deep Fields North and South
using deep Herschel/PACS photometry at 70, 100 and 160 \mics.
Comparing the star-formation rate distributions of X-ray selected 
AGNs to those of inactive galaxies, we confirm that, after accounting for stellar mass selection effects,
AGN hosts lie on the Star-Formation Mass Sequence out to $z\sim2$, consistent with earlier studies \citep{mullaney11b, santini12}. 
However, we also find that AGNs are much more likely to be hosted by a star-forming galaxy
than a quiescent or quenching galaxy. This implies that AGNs are not preferentially found
in galaxies that are undergoing a transformation from the Blue Cloud to the Red Sequence. Instead,
AGN hosts are drawn primarily from the population of normal massive star-forming galaxies.
This may be interpreted in a scenario where radiatively efficient AGNs, such as those selected by X-ray surveys, require
a supply of cold gas to sustain such phases of SMBH accretion. Combined with results which show the lack of any
clear correlation between global SF and nuclear activity in low and moderate luminosity AGNs \citep{shao10, mullaney11b, santini12, rosario12a}, 
our findings suggest that there may not exist any direct causal link between SMBH accretion and overall 
galaxy growth, but rather a more indirect relationship governed by the supply of gas, the fuel for both AGNs
and  star formation.

%However, the vastly different timescales between
%the modulation of AGN activity and galaxy-wide star-formation erases any causal link between SMBH accretion
%and galaxy growth in low and moderate luminosity AGNs.

\acknowledgements{PACS has been developed by a consortium of institutes led by MPE (Germany) and including UVIE
(Austria); KU Leuven, CSL, IMEC (Belgium); CEA, LAM (France); MPIA (Germany); INAF-IFSI/
OAA/OAP/OAT, LENS, SISSA (Italy); IAC (Spain). This development has been supported by the
funding agencies BMVIT (Austria), ESA-PRODEX (Belgium), CEA/CNES (France), DLR (Germany),
ASI/INAF (Italy), and CICYT/MCYT (Spain). FB acknowledges support from Financiamento
Basal, CONICYT-Chile FONDECYT 1101024 and FONDAP-CATA 15010003, and
Chandra X-ray Center grant SAO SP1-12007B (F.E.B.). JRM acknowledges The Leverhulme Trust.
}

\bibliographystyle{apj}

\bibliography{agn_GV_pepgh}

%%%%%%%%%%%%%%%%%%%%%% FIGURE 2 %%%%%%%%%%%%%%%%%%%%%%%%
%\begin{figure*}[t]
%\figurenum{9}
%\centering
%\includegraphics[width=\textwidth]{ssfr-mass_gvcolor_agn.eps}
%\caption
%{ Specfic Star Formation Rate (SSFR) against stellar mass (\smass) of AGNs. Points
%are colored by the \deluv, the $U-V$ offset of the galaxy from the Green Valley (GV), defined by straight lines in the CMD (\S3).
%AGNs that lie in the GV have \deluv\ close to zero and are colored as green points in the Figure.
%PACS-detected AGNs in the GV lie on or around the Mass Sequence (shown as a dashed lines) and are typically massive, 
%while those in the Blue Cloud are typically at lower masses and lie well above the Mass Sequence. 
%For AGNs undetected in PACS, we show estimated upper limits to the FIR luminosity. Both FIR-detected and undetected
%galaxies at high masses can show colors consistent with the Green Valley.
%}
%\end{figure*}
%%%%%%%%%%%%%%%%%%%%%%%%%%%%%%%%%%%%%%%%%%%%%%%%%%

\end{document}